\documentclass[10pt, twocolumn, aps]{revtex4-1} 
\pdfoutput=1	

\usepackage{enumitem} 

\usepackage[ruled]{algorithm}
\usepackage{tabularx}
\usepackage[noend]{algpseudocode}
\makeatletter
\newcommand\fs@nobottomruled{\def\@fs@cfont{\bfseries}\let\@fs@capt\floatc@ruled
  \def\@fs@pre{}
 \def\@fs@post{\kern2pt\hrule\relax}
  \def\@fs@mid{\kern2pt\hrule\kern2pt}
  \let\@fs@iftopcapt\iftrue}
\makeatother
\floatstyle{nobottomruled}
\restylefloat{algorithm}

\makeatletter
\newcommand{\multiline}[1]{%
  \begin{tabularx}{\dimexpr\linewidth-\ALG@thistlm}[t]{@{}X@{}}
    #1
  \end{tabularx}
}
\makeatother
\algrenewcommand\algorithmicindent{1.0em}%

\usepackage{amsfonts, amssymb, amsmath, amsthm}
\theoremstyle{plain}
\usepackage{latexsym}
\usepackage[tracking=smallcaps]{microtype}	
\usepackage{url}
\usepackage{color}
\definecolor{DarkGray}{rgb}{0.1,0.1,0.5}
\usepackage[colorlinks=true,breaklinks, linkcolor=DarkGray,citecolor=DarkGray,urlcolor=DarkGray]{hyperref}	

\usepackage{graphicx}
\usepackage[tight, TABBOTCAP]{subfigure}

\newcommand{\ket}[1]{{|#1\rangle}}

\newcommand{\abs}[1]{{\lvert #1\rvert}}	

\newcounter{sprows}

\newlength{\spheight}
\newlength{\spraise}

\newlength{\commentslength}

\newcommand{\rem}[1]{}

\newtheorem{theorem}{Theorem}
\newtheorem{lemma}[theorem]{Lemma}

\newtheorem*{claim*}{Claim}

\newtheorem{conjecture}[theorem]{Conjecture}

\newtheorem{definition}[theorem]{Definition}

\newfont{\subsubsecfnt}{ptmri8t at 11pt}
\renewcommand{\subparagraph}[1]{\smallskip{\subsubsecfnt #1.}}

\newcommand{\eqnref}[1]{\hyperref[#1]{{(\ref*{#1})}}}
\newcommand{\thmref}[1]{\hyperref[#1]{{Theorem~\ref*{#1}}}}
\newcommand{\lemref}[1]{\hyperref[#1]{{Lemma~\ref*{#1}}}}
\newcommand{\corref}[1]{\hyperref[#1]{{Corollary~\ref*{#1}}}}
\newcommand{\defref}[1]{\hyperref[#1]{{Def.~\ref*{#1}}}}
\newcommand{\secref}[1]{\hyperref[#1]{{Sec.~\ref*{#1}}}}
\newcommand{\figref}[1]{\hyperref[#1]{{Fig.~\ref*{#1}}}}  
\newcommand{\tabref}[1]{\hyperref[#1]{{Table~\ref*{#1}}}}
\newcommand{\remref}[1]{\hyperref[#1]{{Remark~\ref*{#1}}}}
\newcommand{\appref}[1]{\hyperref[#1]{{Appendix~\ref*{#1}}}}
\newcommand{\claimref}[1]{\hyperref[#1]{{Claim~\ref*{#1}}}}
\newcommand{\factref}[1]{\hyperref[#1]{{Fact~\ref*{#1}}}}
\newcommand{\propref}[1]{\hyperref[#1]{{Proposition~\ref*{#1}}}}
\newcommand{\exampleref}[1]{\hyperref[#1]{{Example~\ref*{#1}}}}
\newcommand{\conjref}[1]{\hyperref[#1]{{Conjecture~\ref*{#1}}}}

\allowdisplaybreaks[1]

\def\COLOR{}
\ifdefined\COLOR

\else

\fi

\definecolor{Cayenne}{rgb}{0.5,0,0}
\definecolor{Midnight}{rgb}{0,0,0.5}
\definecolor{Plum}{rgb}{0.5,0,0.5}
\definecolor{Teal}{rgb}{0,0.5,0.5}
\definecolor{Clover}{rgb}{0,0.5,0}
\definecolor{Maroon}{rgb}{0.5,0,0.25}
\definecolor{Ocean}{rgb}{0,0.25,0.5}
\definecolor{Tangerine}{rgb}{1,0.5,0}
\definecolor{Strawberry}{rgb}{1,0,0.5}
\definecolor{Fern}{rgb}{0.25,0.5,0}
\definecolor{Aqua}{rgb}{0,0.5,1}
\definecolor{Moss}{rgb}{0,0.5,0.25}
\definecolor{Mocha}{rgb}{0.5,0.25,0}
\definecolor{Lemon}{rgb}{1,1,0}
\definecolor{Asparagus}{rgb}{0.5,0.5,0}
\definecolor{Grape}{rgb}{0.5,0,1}
\definecolor{Iron}{rgb}{.3,.3,.3}
\definecolor{Steel}{rgb}{.4,.4,.4}

\usepackage{changepage} 

\usepackage{enumitem} 

\makeatletter
\let\save@mathaccent\mathaccent
\newcommand*\if@single[3]{%
  \setbox0\hbox{${\mathaccent"0362{#1}}^H$}%
  \setbox2\hbox{${\mathaccent"0362{\kern0pt#1}}^H$}%
  \ifdim\ht0=\ht2 #3\else #2\fi
  }
\newcommand*\rel@kern[1]{\kern#1\dimexpr\macc@kerna}
\newcommand*\widebar[1]{\@ifnextchar^{{\wide@bar{#1}{0}}}{\wide@bar{#1}{1}}}
\newcommand*\wide@bar[2]{\if@single{#1}{\wide@bar@{#1}{#2}{1}}{\wide@bar@{#1}{#2}{2}}}
\newcommand*\wide@bar@[3]{%
  \begingroup
  \def\mathaccent##1##2{%
    \let\mathaccent\save@mathaccent
    \if#32 \let\macc@nucleus\first@char \fi
    \setbox\z@\hbox{$\macc@style{\macc@nucleus}_{}$}%
    \setbox\tw@\hbox{$\macc@style{\macc@nucleus}{}_{}$}%
    \dimen@\wd\tw@
    \advance\dimen@-\wd\z@
    \divide\dimen@ 3
    \@tempdima\wd\tw@
    \advance\@tempdima-\scriptspace
    \divide\@tempdima 10
    \advance\dimen@-\@tempdima
    \ifdim\dimen@>\z@ \dimen@0pt\fi
    \rel@kern{0.6}\kern-\dimen@
    \if#31
      \overline{\rel@kern{-0.6}\kern\dimen@\macc@nucleus\rel@kern{0.4}\kern\dimen@}%
      \advance\dimen@0.4\dimexpr\macc@kerna
      \let\final@kern#2%
      \ifdim\dimen@<\z@ \let\final@kern1\fi
      \if\final@kern1 \kern-\dimen@\fi
    \else
      \overline{\rel@kern{-0.6}\kern\dimen@#1}%
    \fi
  }%
  \macc@depth\@ne
  \let\math@bgroup\@empty \let\math@egroup\macc@set@skewchar
  \mathsurround\z@ \frozen@everymath{\mathgroup\macc@group\relax}%
  \macc@set@skewchar\relax
  \let\mathaccentV\macc@nested@a
  \if#31
    \macc@nested@a\relax111{#1}%
  \else
    \def\gobble@till@marker##1\endmarker{}%
    \futurelet\first@char\gobble@till@marker#1\endmarker
    \ifcat\noexpand\first@char A\else
      \def\first@char{}%
    \fi
    \macc@nested@a\relax111{\first@char}%
  \fi
  \endgroup
}
\makeatother

\def\beq{\begin{equation}}
\def\eeq{\end{equation}}

\begin{document}

\title{Flag fault-tolerant error correction for any stabilizer code}
\author{Rui Chao}
\author{Ben W.\ Reichardt}
\affiliation{University of Southern California}

\begin{abstract} 
Conventional fault-tolerant quantum error-correction schemes require a number of extra qubits that grows linearly with the code's maximum stabilizer generator weight.  For some common distance-three codes, the recent ``flag paradigm" uses just two extra qubits.  Chamberland and Beverland (2018) provide a framework for flag error correction of arbitrary-distance codes.  However, their construction requires conditions that only some code families are known to satisfy. 

We give a flag error-correction scheme that works for any stabilizer code, unconditionally.  With fast qubit measurement and reset, it uses $d+1$ extra qubits for a distance-$d$ code.  
\end{abstract}

\maketitle

\section{Introduction}

In quantum error correction, errors are diagnosed by measuring the code's check operators. 
The measurement circuits themselves are faulty. 
One way to measure the syndromes fault tolerantly is to prepare some extra qubits in a special state, then couple these ancillas to the data qubits. 
Conventional fault-tolerant error-correction schemes~\cite{Shor96, Steane97, Knill03erasure, DiVincenzoAliferis06slow} need as many ancillas as the maximum stabilizer generator weight. 
The ``flag method''~\cite{Chao2018ec,Chao2018compute}, however, requires only two ancilla qubits for common distance-three codes. 
Chamberland and Beverland~\cite{ChamberlandBeverland17flags} have generalized the flag idea to arbitrary-distance codes. 
Here, we give an explicit, ancilla-efficient flag fault-tolerant error-correction scheme that unconditionally applies to arbitrary stabilizer codes. 

\begin{figure}[!b]
\includegraphics[scale=.75]{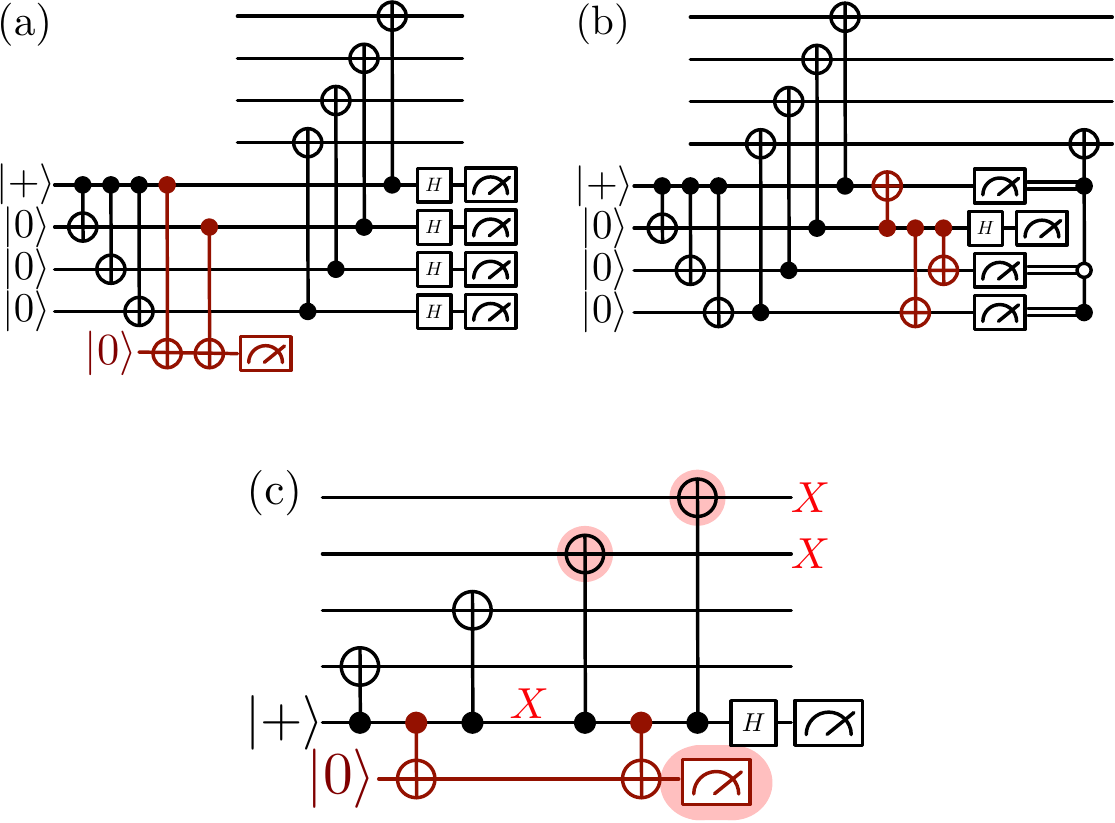}
\caption{
Different fault-tolerant methods for measuring~$X^{\otimes4}$. 
(a) Shor's method. 
Prepare a four-qubit cat state and then detect correlated errors using an extra ancilla qubit (red). 
Apply subsequent transversal gates to the data only when the detection  outcome is trivial. 
(b) DiVincenzo-Aliferis decoding method.  
Couple the data qubits with an unverified cat state. 
Then apply the decoding circuit (red) followed by qubit-wise measurements, the outcomes of which give the syndrome and information about potential correlated data errors. 
This method has been extended to measure an arbitrary-weight stabilizer of a distance-three code~\cite[Supp.~Mat.]{Chao2018ec}. 
(c)~Flag method. 
Extract the syndrome into an ancilla qubit, which is protected by a flag gadget (red). 
A single fault, e.g., $X$, on the syndrome ancilla can propagate into  correlated data errors, e.g., $X^{\otimes 2}$. 
Such correlated errors are signaled by a nontrivial measurement outcome on the flag ancilla, and can be diagnosed using subsequent syndrome measurements. 
}\label{f:threeschemes}
\end{figure}

\hyperref[f:threeschemes]{{Figure~1}} compares the flag method to the Shor~\cite{Shor96} and DiVincenzo-Aliferis decoding~\cite{DiVincenzoAliferis06slow} methods.   
In the flag method, the syndrome is extracted into a single ancilla, on which one gate failure can spread to two or more errors on the data. 
To fix this problem, the syndrome ancilla is protected with a flag gadget.  
The flag measurement signals the possible occurrence of any correlated errors.
Provided that the flag circuit is carefully designed, the possible errors are distinguishable by their syndromes and so can be corrected. 

Flag error-correction schemes have been given for several code families: Hamming codes~\cite{Chao2018ec}, rotated surface and Reed-Muller codes~\cite{ChamberlandBeverland17flags}, color codes~\cite{ChamberlandBeverland17flags,reichardt18steane, baireuther19nnd, bermudez19memory, chamberland19color},  cyclic CSS codes~\cite{tansuwannont2018flag}, and heavy hexagon and square codes~\cite{chamberland19heavy}.  
Flag-based schemes have been proposed also for error detection~\cite{vuillot17ibm}, logical state preparation~\cite{chamberland19magicpreparation, shi19GKP} and lattice surgery~\cite{gutierrez19surgery, lao19bridge}.
In particular, Chamberland and Beverland~\cite{ChamberlandBeverland17flags} have formalized the flag idea for error correction and extended it to arbitrary-distance codes. 
For a distance-$d$ code, they have shown how to construct a flag error-correction protocol, provided that the ``detectability'' and ``distinguishability'' requirements are satisfied. 
(Detectability requires that each stabilizer generator is measured with a flag circuit that signals whenever $s\le\left\lfloor\frac{d-1}{2}\right\rfloor$ faults result in more than $s$ data errors. 
Distinguishability requires that any two data error patterns, each arising from at most $\left\lfloor\frac{d-1}{2}\right\rfloor$ faults spread among the stabilizer generators' flag circuits and resulting in same flag pattern, either have distinct syndromes or differ by a stabilizer.)

Here, we provide a flag error-correction scheme that applies to any stabilizer code, unconditionally.  
In particular, the circuit construction relies only on the code's distance and maximum stabilizer generator weight.
The key idea is that harmful faults can be approximately located from the observed flag pattern and then corrected right away, before they propagate to correlated data errors. 
For a distance-$d$ code, the scheme requires one syndrome ancilla and $d$ flag ancillas, $d+1$ extra qubits total. 
No further qubits are needed, provided that the ancillas can be reset rapidly, in time comparable to a gate operation.  

Different fault-tolerant error-correction schemes are summarized in~\tabref{t:summary}. 
Unlike Shor's and similar schemes, our procedure does not verify the ancillas; no postselection is needed. 
Unlike previous flag schemes, ours is non-adaptive, in the sense that the flag measurement outcomes alone provide enough information to correct correlated errors; no extra syndrome measurement is required. 
Prabhu and Reichardt~\cite{PrabhuReichardt} have recently provided new flag error-correction schemes that apply to arbitrary distance-three and -five codes, and require, respectively, $O(\log w)$ and $O(w)$ ancilla qubits, where $w$ is the maximum stabilizer generator weight. 
Their schemes are also non-adaptive but allow slow qubit reset; each ancilla qubit is measured only once at the end. 

\begin{table}
\begin{center}
\begin{tabular}{|c|@{\hskip0pt}c@{\hskip0pt}|@{\hskip1pt}c@{\hskip0pt}|}
\hline \hline
& Adaptive & Non-adaptive \\ 
\hline
Slow &  & \hspace{.3cm} $\lceil w/2\rceil$ \hspace{.36cm} for $d=3$~\cite{Chao2018ec}  \\
qubit & None & \hspace{.1cm} $O(\log w)$ \hspace{.38cm} for $d=3$~\cite{PrabhuReichardt} \\
reset & &  $\lceil w/2\rceil+O(1)$ for $d=5$~\cite{PrabhuReichardt}  \\
\hline
Fast & $O(1)$ for many codes & $\Omega(w)$~\cite{Shor96,Steane97,Knill03erasure} \\
reset &\cite{Chao2018ec,ChamberlandBeverland17flags,tansuwannont2018flag} & $d+1$ for all $d$ (\secref{s:fastd}) \\
\hline \hline
\end{tabular}
\end{center}
\caption{
Numbers of ancilla qubits for different error-correction schemes. 
Here, $d$ is the code distance and $w$ is the maximum stabilizer generator weight. 
In~\cite{Chao2018ec,ChamberlandBeverland17flags,tansuwannont2018flag}, flag patterns can only detect or reveal partial information about the correlated data errors; adaptive syndrome measurement is further needed for full correction. 
In~\cite{Shor96,Steane97,Knill03erasure}, correlated data errors cannot occur due to ancilla verification. 
In the other schemes in the table, syndrome-measurement circuits can be scheduled deterministically; (flag) ancilla patterns provide enough information to correct correlated errors.
} \label{t:summary}
\end{table}

\section{Fault-tolerance definitions}
\label{s:definition}
We restrict to codes with odd distance $d=2t+1$. 
Faults can occur in elementary operations: single-qubit preparations and measurements, single- and two-qubit gates, and idle positions.  

\begin{definition}[\cite{AliferisGottesmanPreskill05,Gottesman09faulttolerance}]\label{d:ft}
An error-correction protocol is $d$-fault-tolerant ($d$-FT) if it satisfies
\begin{enumerate}[leftmargin=*]
\item If the input has $r$ errors and the protocol has $s$ faults with $r+s\le t$, then the output is correctable by perfect decoding. 
\item If the protocol has $s$ faults with $s\le t$, then the output is at most $s$ errors away from some codeword. 
\end{enumerate}
\end{definition}

Our protocol measures one stabilizer generator at a time, like the schemes in \figref{f:threeschemes}.  
The \emph{flag circuit} uses one \emph{syndrome ancilla} and one or more \emph{flag ancillas}.  If there are no faults, measuring the syndrome ancilla gives the syndrome, and the flag ancilla measurement outcomes will be trivial. 

\begin{definition} \label{d:flagft}
A flag syndrome-measurement circuit is $d$-FT if any $k\le t$ faults result in at most $k$ data errors, after corrections based on the flag pattern.  
\end{definition}

For example, the circuit in~\hyperref[f:threeschemes]{{Fig.~1(b)}} is 3-FT for a weight-four stabilizer. 
An important observation is that one can construct a $d$-FT error-correction protocol by measuring one stabilizer generator at a time, each using a $d$-FT flag circuit.  
Indeed,   $d$-FT circuits automatically guarantee condition 1 of~\defref{d:ft}. 
One way to achieve condition 2 is to keep measuring the syndromes until the same syndrome repeats consecutively a certain number of times without flags being triggered~\cite{Shor96}. 

Notice that \defref{d:flagft} is stronger than that of a \emph{$t$-flag circuit}, introduced by~\cite{ChamberlandBeverland17flags}, which guarantees a nontrivial flag pattern whenever $s\le t$ faults result in more than $s$ errors.   
(In particular, \cite{ChamberlandBeverland17flags} gives a scheme for constructing a 2-flag circuit that requires five ancilla qubits; and gives a conjectured scheme for constructing a general $t$-flag circuit that requires $w$ ancilla qubits.)

Denote by $|f|$ the Hamming weight, and by $f\oplus g$ the bitwise XOR, of binary strings $f$ and $g$. 
Single-qubit measurements in diagrams are all in the $Z$ basis ($\ket0,\ket1$).

\section{Syndrome measurement for \\any stabilizer code} \label{s:fastd}

In this section, we present a flag circuit that measures a stabilizer of an arbitrary distance-$d$ code. 
Without loss of generality, we may take the stabilizer to be $X^{\otimes w}$. 

\begin{figure}[!h]
\centering
\begin{tabular}{cc}
\subfigure[\label{f:fastflagsd3w4}]{
\hspace{-1.2cm}\raisebox{-1.73cm}{\includegraphics[scale=.75]{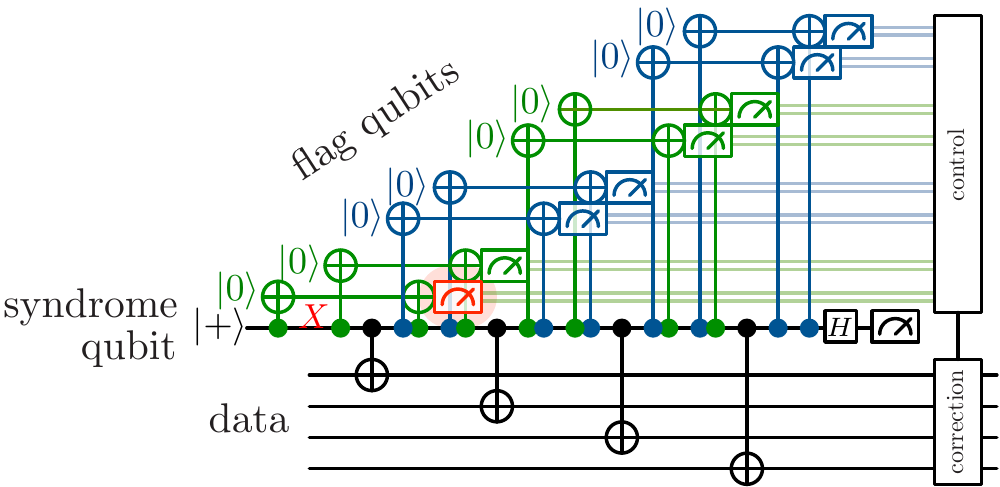}}
}\\ 
\hspace{-.1cm}\subfigure[\label{f:fastflagsschematic}]{
\hspace{-.66cm}\raisebox{-1.73cm}{\includegraphics[scale=.66]{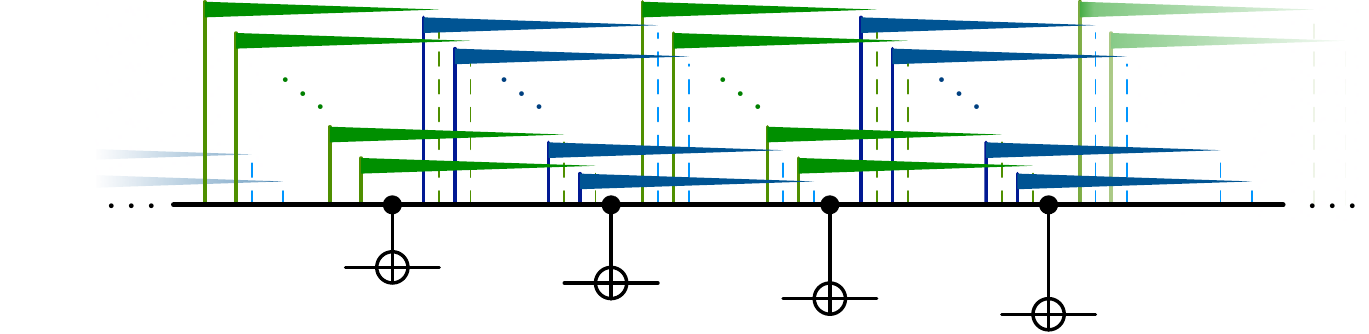}}
}
\end{tabular}
\caption{
(a) Distance-three fault-tolerant measurement of~$X^{\otimes 4}$.  Each CNOT to the data is protected by two flags. $X$ corrections are applied based on the flag pattern.   
(b)~Schematic showing how the circuit extends to $X^{\otimes w}$ and arbitrary distance~$d$; control and corrections are omitted. 
Between each consecutive pair of CNOT gates to the data, $d - 1$ flags are turned on, and $d - 1$ turned off, in an alternating fashion: on, off, on, off.  
} \label{f:fastflagsarbitraryd}
\end{figure}

\hyperref[f:fastflagsarbitraryd]{{Figure~2}} illustrates the construction.  
A syndrome ancilla, initialized in $\ket +$, is coupled to the $w$ data qubits with CNOT gates.  Each of these CNOT gates is protected by a different set of $d - 1$ flags.  Between CNOT gates to the data, the flags are carefully transitioned---with flag~$j$ for the next CNOT turned on just before flag~$j$ for the previous CNOT is turned off, for $j \in [d - 1]$---so that there are always either $d - 1$ or $d$ active flags, except at the very beginning and end.  
Observe that with fast qubit reset, this construction requires at most $d$ flag ancillas, so $d + 1$ ancilla qubits total. 

\begin{conjecture} \label{t:flagconjecture}
With appropriate correction rules based on the flag pattern, the procedure of \figref{f:fastflagsarbitraryd} is $d$-FT.  
\end{conjecture}

\begin{figure}
\centering
\includegraphics[scale=.8]{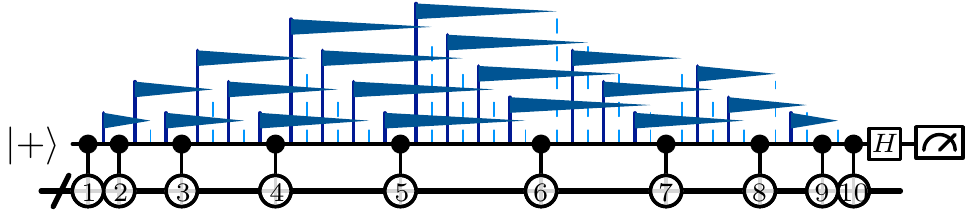}
\caption{An optimized circuit for measuring $X^{\otimes 10}$.  
Note that the central data CNOT gates are protected by $d - 1$ flags, but fewer flags are used toward the ends.  
With appropriate correction rules, the circuit is fault tolerant to distance $d = 5$.  
} \label{f:fastflagsd5w10}
\end{figure}

A stronger statement might also hold.  
Based on computer analyses of circuits for $d \in \{3, 5, 7\}$ (see~\appref{a:pseudocode}), it appears that the data CNOTs toward the beginning or end require fewer flags: $0, 1, 2, \ldots, {d-2}, {d-1}, {d-1}, \ldots, {d-1}, {d-2}, \ldots, 2, 1, 0$ flags instead of always~$d-1$ flags.  
The first and last data CNOTs require no flag protection because a syndrome-ancilla fault immediately before or after these gates can propagate to an error of weight only $0$ or $1$.  Moving toward the middle, the number of flags needed then steadily increases up to a maximum of $d - 1$.  \hyperref[f:fastflagsd5w10]{{Figure~3}}  shows an example, with $w = 10$ and $d = 5$.    

However, we will not attempt to prove \conjref{t:flagconjecture}.  Instead, in order to simplify the correction rules' boundary conditions, modify the scheme of \figref{f:fastflagsarbitraryd} by prepending $r = (d + 1)^2 / 4$ rounds of $d - 1$ flags.  
That is, follow \figref{f:fastflagsarbitraryd} for a weight $r + w$ stabilizer, except do not use data CNOTs in the first $r$ rounds.  
We will show: 

\begin{lemma} \label{c:mainthm}
The modified procedure, with $(d + 1)^2 / 4$ initial flag rounds and appropriate correction rules, is $d$-FT.  
\end{lemma}

\begin{figure}
\centering
\raisebox{-1.73cm}{\includegraphics[scale=.8]{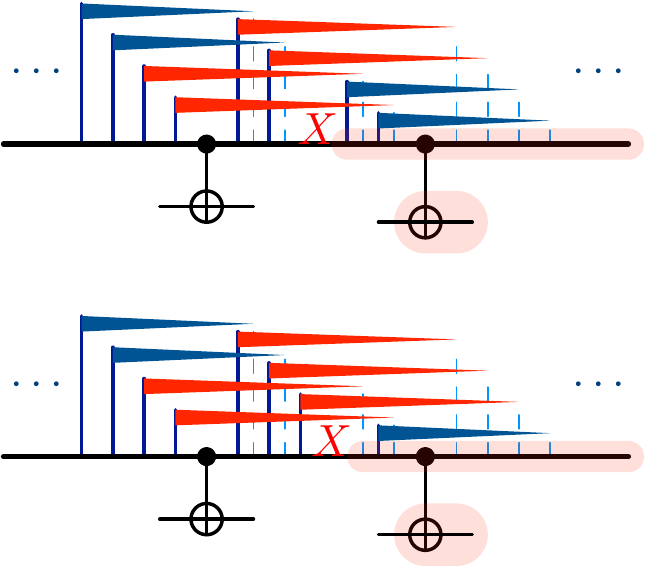}}
\caption{
A syndrome fault can flip $d - 1$ or $d$ flags.  
} \label{f:faultflags}
\end{figure}

\begin{figure}
\centering
\includegraphics[scale=.73]{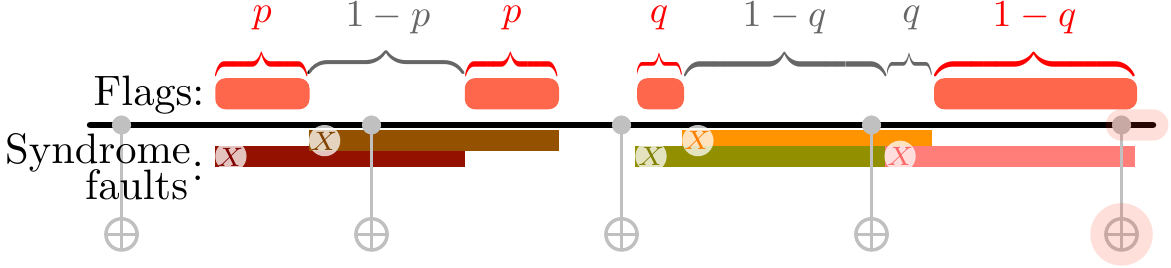}
\caption{Syndrome faults spread to the data and flip intervals of flags.  The correction algorithm uses the observed flag pattern, which can be further distorted by flag faults, to infer the syndrome faults' approximate locations.  
} \label{f:flagintervals}
\end{figure}

Let us first give some intuition.  
Distinguish possible circuit faults into syndrome faults and flag faults. 
Syndrome faults cause $X$ faults on the syndrome ancilla; they can spread to correlated data errors. 
Flag faults are measurement faults on flag ancillas; they do not spread to the data. 
A syndrome fault can trigger $d-1$ or $d$ flags, as in \figref{f:faultflags}, or $< d-1$ flags on the left and right boundaries, as in \figref{f:fastflagsd3w4}.  
The motivation for the initial flag rounds is so that our correction algorithm, which works from left to right, does not need to account for left boundary conditions.  

Assume for the moment that every syndrome fault flips exactly $d - 1$ flags.  
Imagine laying the flags along the real line, so that successive flags cover consecutive intervals of length $1/(d - 1)$; then each syndrome fault flips an interval of length one. 
The up to $t$ flag faults can further flip intervals with total length up to $\tfrac{t}{d-1} = 1/2$.  

As shown in \figref{f:flagintervals}, the pattern of triggered flags depends on how the flipped intervals overlap.  The correction algorithm needs to use the observed flags to deduce the approximate locations of the syndrome faults.  It does not need to determine the exact locations because all that matters is how the faults propagate to data-qubit errors, via the upward data CNOTs located at integers on the line.  Also, an even number of syndrome faults with overlapping intervals, as on the left side of \figref{f:flagintervals}, does not require any data correction for fault tolerance.  

\begin{proof}[Proof of \lemref{c:mainthm}]
We will give an explicit correction algorithm and then analyze it inductively.  

Since a syndrome fault can trigger flags in up to two consecutive flag rounds, with at least $t (t-1+2) + t = t (t + 2)$ initial flag rounds, there must be at least one sequence of at least $t$ consecutive flag rounds in which no flags are triggered.  Call this sequence the ``start" rounds.  
Furthermore, the correction algorithm can identify start rounds without knowing the fault locations.  (Although faults can cancel out, they cannot over such a long stretch.  As on the right side of \figref{f:flagintervals}, three syndrome faults cancel out over a length-one interval, more generally $t$ faults can cancel out an interval of length at most $t / 2$.)  Having identified the start rounds, the algorithm discards all flags to their left.  
(Faults on the syndrome ancilla before it interacts with the data do not cause any data errors.)  

For fault tolerance, it suffices to consider only syndrome and flag faults; any other fault either can be ignored or is equivalent to a syndrome or flag fault.  Divide the faults after the start rounds into three sets: set $F$ of flag faults, set $S_d$ of syndrome faults flipping exactly $d$ flags, and set $S_{d-1}$ of syndrome faults either flipping $d-1$ flags or on the last round.  
They satisfy $|F|+|S_d|+|S_{d-1}|\le t$. 
For instance, a $YY$ fault on a data CNOT is in $S_{d-1}$.  

\smallskip

Now we specify our correction algorithm, and then prove that syndrome measurement is $d$-fault tolerant.  

\smallskip

The algorithm will work through the flags from the start rounds on, from left to right, interpreting one round of $d - 1$ flags at a time.  
Denote by $f_1,f_2,\ldots \in \{0,1\}^{2t}$ the rounds of flags after the start rounds, $f_\ell$ being the flags around data (or dummy) CNOT~$\ell$, as in \figref{f:induction}. 
Let $\Omega$ be the rightmost nontrivial round, and define $\varepsilon\in\{0,1\}^{2t},|\varepsilon|=1$ as the rightmost nontrivial flag of~$f_\Omega$.  

\begin{figure}[!h]
\centering
\includegraphics[scale=.5]{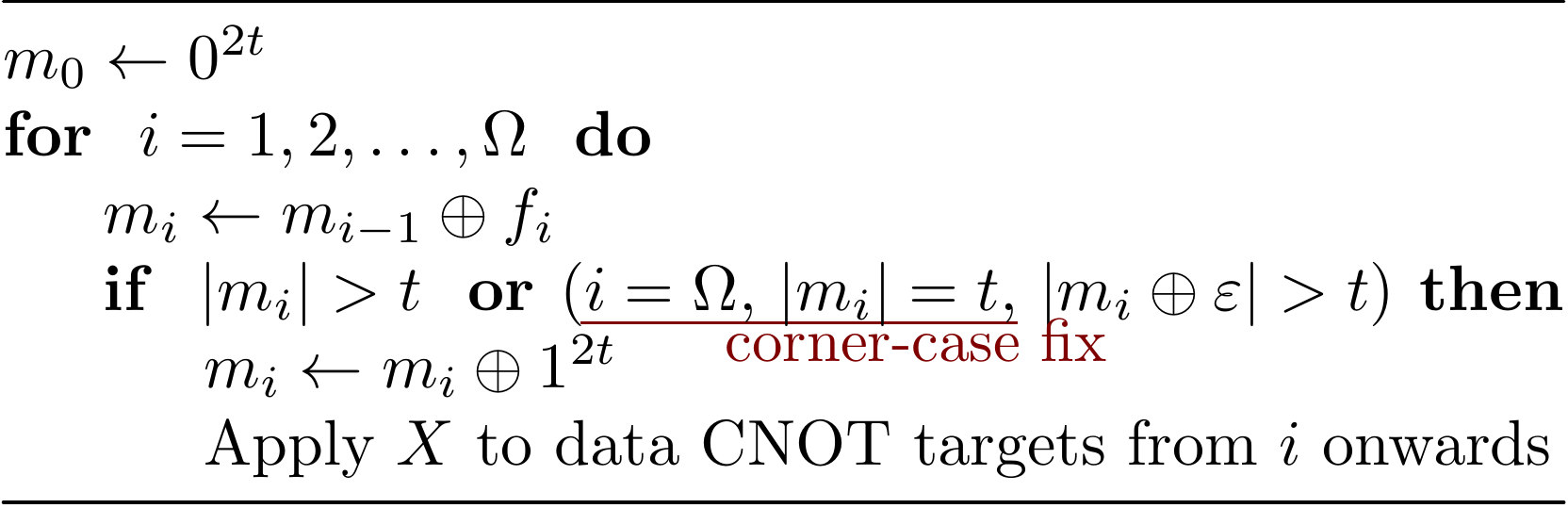}
\end{figure}

The algorithm deals with each round in the same way except round $\Omega$.
When $i=\Omega$ and $|m_i|=t$, the corner-case fix effectively flips the rightmost nontrivial flag $\varepsilon$. 

\smallskip 

The proof proceeds by incorporating faults one at a time but with $\Omega$ and $\varepsilon$ fixed.  
We compare the algorithm's behavior before and after each fault is incorporated.  
For fault tolerance, we argue that the residual data errors after the corrections never outnumber the current faults.  

\def\newF{{\overline F}}

Split each fault in $S_d$ into a $d-1$ bit syndrome fault, plus one extra flag fault at the end.  That is, define $\newF$ to be the union of $F$ and the rightmost bits from $S_d$, and let $S$ be the union of $S_{d-1}$ and remaining parts of $S_d$.  (Thus, $|\newF|=|F|+|S_d|,|S|=|S_{d-1}|+|S_d|$.)  

For the base case, we consider all the faults in $\newF$ with no data errors. 
Define $r_1, \ldots, r_\Omega \in \{0,1\}^{2t}$ as the flag patterns flipped by faults in $\newF$.  
In subsequent inductive steps, we add faults in $S$, from left to right, one at a time.  Suppose the added fault $s$ occurs between data CNOTs $\ell$ and~$\ell+1$, causing data errors $P_\ell X_{\ell+1}X_{\ell+2}\cdots$ for some $P \in \{I,X,Y,Z\}$. 
We will prove, by induction in $s$, that for all $i \geq \ell+1$, the value of $m_i$ when the algorithm finishes satisfies $m_i=\bigoplus_{j=1}^{i}r_j$.  
(For the base case, set~$\ell = 0$.)  

\smallskip

\begin{figure}
\centering
\hspace{-.17cm}\includegraphics[scale=.75]{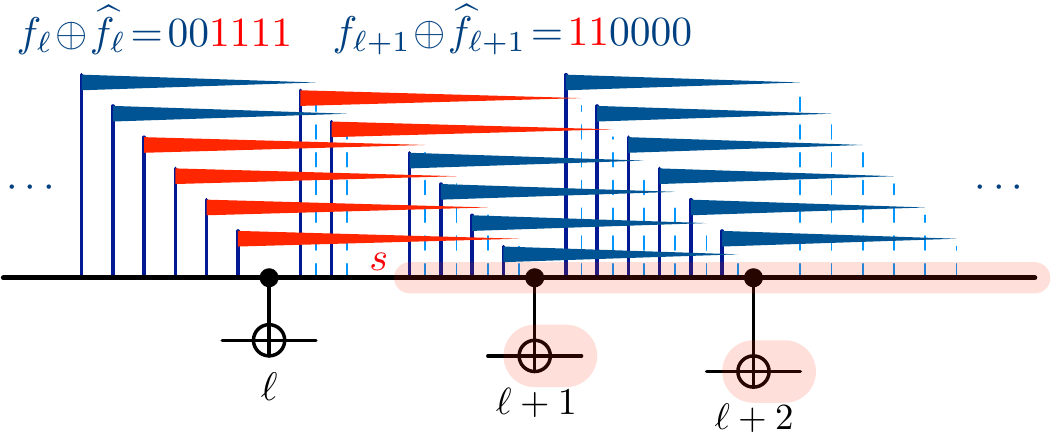}
\caption{
If $d=7$, $s\in S$ flips six flags in $f_\ell$ and $f_{\ell+1}$.  
} \label{f:induction}
\end{figure}

Consider the base case of faults in $\newF$.  Then $f_i=r_i$. 
We claim that $m_i=\bigoplus_{j=1}^{i} r_j$.  This is clear since $\big|\!\bigoplus_{j=1}^i r_j \big| \le \abs{\newF} \leq t$ and so corrections are never triggered except possibly at round $\Omega$.  
If $\left|m_{\Omega-1}\oplus r_\Omega \right|=t$, then necessarily $S_{d-1} = \varnothing$ and the flag $\varepsilon$ is attributed to the rightmost fault in $\newF$. 
In particular, $|m_\Omega\oplus\varepsilon|=t-1$. 
Therefore, $|m_i|$ or $|m_\Omega\oplus\varepsilon|$ never exceeds $t$, and the algorithm applies no corrections, as desired. 

\smallskip

Now consider adding a syndrome fault $s\in S$, between data CNOTs $\ell$ and $\ell+1$. 
Define $\widehat f_i$ and $ \widehat  m_i$ as the parameters of the algorithm (when finished) in the previous induction step, without~$s$. 
(See~\figref{f:induction} for an example.) 
With $s$ added, the algorithm's behavior does not change until~$i = \ell$. 
(Recall that $\Omega$ and $\varepsilon$ are fixed.)
Beyond $\ell$, the following observation will essentially finish the proof. 

\smallskip

\noindent
\textbf{Observation.} In exactly one of the  two rounds $i=\ell$ and $\ell+1$ the algorithm changes its behavior: either applying the correction when it did not before, or vice versa. 
The algorithm's behavior remains unchanged for $i \geq \ell+2$. 

Since $s$ flips $2t$ flags, we have
\begin{equation} \label{e:flagflip}
\big(\widehat f_{\ell}\oplus f_\ell \big) \oplus \big( \widehat f_{\ell+1} \oplus f_{\ell+1} \big)=1^{2t} \enspace .
\end{equation}
Note that $m_\ell \oplus m_{\ell-1} \oplus f_\ell \in \{1^{2t}, 0^{2t}\}$, depending on whether in round~$\ell$, $X$ data corrections are applied or~not.  
To show the observation, we consider all possible scenarios.  

\begin{enumerate}[leftmargin=*]
\item $\widehat m_\ell \oplus \widehat m_{\ell-1} \oplus \widehat f_\ell \neq  m_\ell \oplus  m_{\ell-1} \oplus  f_\ell $, i.e., the correction behavior changes at round $\ell$. 
Since $\widehat m_{\ell-1}=m_{\ell-1}$, we have $\widehat m_{\ell}\oplus\widehat f_{\ell} \oplus m_\ell\oplus f_\ell = 1^{2t}$. 
Adding \hyperref[e:flagflip]{{Eq.~(1)}}, $m_\ell\oplus f_{\ell+1}=\widehat m_\ell\oplus \widehat f_{\ell+1}$, and the algorithm's behavior does not change from round $\ell+1$ on. 
In particular, for $i\ge\ell+1$, still $m_i = \widehat m_i = \bigoplus_{j=1}^i r_j$. 
The change of corrections $X_\ell X_{\ell+1} \cdots$ leaves at most one data error $P_\ell X_\ell$, as desired. 

\item $\widehat m_\ell \oplus \widehat m_{\ell-1} \oplus \widehat f_\ell =  m_\ell \oplus  m_{\ell-1} \oplus  f_\ell $, i.e., the correction behavior does not change at round $\ell$. 
Adding $\widehat m_{\ell-1} = m_{\ell-1}$ and \hyperref[e:flagflip]{{Eq.~(1)}} gives $\big( \widehat m_\ell \oplus \widehat f_{\ell+1} \big) \oplus ( m_\ell \oplus f_{\ell+1} ) = 1^{2t}$. 
We claim the correction behavior changes at round $\ell+1$, and so the change of corrections $X_{\ell+1}X_{\ell+2}\cdots$ leaves at most one data error $P_\ell$, as desired.

\vspace{-.2cm}
\begin{enumerate}[leftmargin=.4cm]
\item $\big| \widehat m_\ell \oplus \widehat f_{\ell+1} \big| \neq t$. 
Then one of the terms $\widehat m_\ell \oplus \widehat f_{\ell+1}$ and $m_\ell \oplus f_{\ell+1}$ has weight $> t$ and the other $< t$. 
The correction behavior changes in round $\ell+1$. 
Thus, we have $\widehat m_{\ell+1} \oplus m_{\ell+1} = (\widehat m_\ell \oplus \widehat f_{\ell+1}) \oplus (m_\ell \oplus f_{\ell+1}) \oplus 1^{2t} = 0^{2t}$ and so $m_i = \widehat m_i = \bigoplus_{j=1}^i r_j$ for $i\ge\ell+1$. 

\vspace{-.1cm}
\item $\big| \widehat m_\ell \oplus \widehat f_{\ell+1} \big| = t$. 
Then $\abs{\widehat m_{\ell+1}} = t$.  
Then since $\widehat m_{\ell+1} = \bigoplus_{j=1}^{\ell+1} r_j$ and $\big| \bigoplus_{j=1}^{\ell+1} r_j \big| \le |\newF|$, we have $S_{d-1} = \varnothing$. 
Therefore, $\ell+1 = \Omega$ and the corner-case fix is triggered in round $\ell+1$, both with and without~$s$. Since $\big( \widehat m_\ell \oplus \widehat f_{\ell+1} \big) \oplus ( m_\ell \oplus f_{\ell+1} ) = 1^{2t}$ and $|\varepsilon|=1$, $\big\{\left| \varepsilon\oplus m_\ell  \oplus f_{\ell+1} \right| , \big| \varepsilon \oplus \widehat m_\ell \oplus \widehat f_{\ell+1} \big| \big\} = \{t-1, t+1\}$. 
As one of the terms has weight $> t$, the correction behavior changes in round $\ell+1$. 
\end{enumerate}
\end{enumerate}
\vspace{-.2cm}
When $s$ occurs after the last data CNOT, it can flip fewer than $2t$ flags---but since it causes no data errors, it can be ignored. 
\end{proof}

Note that when the algorithm finishes $|m_\Omega|$ is a lower bound of the number of faults that have occurred in the  syndrome-measurement circuit. 
This information, combined with the measured syndrome, can be used to reduce the number of repetitions of syndrome measurements in a $d$-FT error-correction protocol~\cite{Shor96,Zalka97,ChamberlandBeverland17flags}. 

\smallskip

Intuitively, it seems that our protocol might be optimal in the required number of ancilla qubits.  
Consider a general flag error-correction protocol that applies to arbitrary stabilizer codes.  
In particular, if the number of ancillas consumed is independent of the stabilizer weight~$w$, then when $w\gg d$ there exists an ancilla $\alpha$ coupled via CNOTs with a subset of $\gg d$ relevant data qubits.  
A fault on $\alpha$ can result in disastrous correlated errors. 
Hence the flags triggered by such a fault, corrupted by up to $t-1$ flag faults, should be distinguishable from a flag pattern with $t$ flag faults. 
Thus $\alpha$ should be protected by at least $(t-1) + t + 1 = d - 1$ flags. 
The number of flags protecting $\alpha$ changes by one at a time, so in order to keep at least $d-1$ flags on it, at some time it is protected by $d$ flags.  
In total, counting $\alpha$, at least $d+1$ ancillas are needed.  

\smallskip

An open problem is to reduce or even remove the initial flag rounds while remaining $d$-FT, as in \conjref{t:flagconjecture}. 
This might require drastically different correction rules.  
It would also be interesting to optimize our circuit construction and correction rules for specific code families, and qubit geometry and connectivity constraints.  

\medskip
Research supported by NSF grant CCF-1254119, ARO grant W911NF-12-1-0541, and MURI Grant FA9550-18-1-0161.

\appendix

\section{Computer search for correction rules}
\label{a:pseudocode}

Given an arbitrary flag circuit that measures a stabilizer generator of a specific code, it is generally difficult to tell whether there exist correction rules to achieve fault tolerance. 
The pseudocode below runs a brute-force search for $d$-FT correction rules, given a flag circuit $C$ as input.  
Again, the stabilizer is taken to be $X^{\otimes w}$, without loss of generality. 

In the pseudocode, $\mathtt{flag}$ and $\mathtt{error}$ denote the flag pattern and data error pattern resulting from a set of circuit faults, respectively. 

Recall that $C$ consists of only one syndrome ancilla, possibly multiple flag ancillas, and only CNOT gates. 
Any relevant fault in $C$ is equivalent to some element of~$\mathcal F$, except for two-qubit faults on data CNOTs, with $X$ on the control and $P\in\{I,X,Y,Z\}$ on the target---we account for them with the subset $\mathcal F^*$.

$\mathcal D$ is a dictionary whose keys are possible flag patterns generated by combinations of at most $t$ faults. 
For each key, we collect all possible afflicting fault combinations. 
There exist $d$-FT correction rules if and only if for each key-value pair $(\mathtt k,\mathtt v)\in\mathcal D$ there exist consistent corrections for fault combinations in $\mathtt v$. 

Specifically, for each $F\in\mathtt v$, $\mathcal S_F$ contains all the valid $X$ data corrections for $F$, in form of binary strings. 
Indeed, $\overline{\mathcal Q}=[w]\setminus\mathcal Q$ contains all the data qubits whose errors are unknown Paulis, thus corrections on $\overline{\mathcal Q}$ can be arbitrary. 
However, corrections on $\mathcal Q$ should remove sufficient errors so that the residual data error $E_{\mathcal Q}$ has weight no greater than $|F\setminus\mathcal F^*|$. 

Conditioned on success, the code returns a set $\mathcal R$ of valid correction rules indexed by the observed flag pattern.

\begin{figure}[!h]
\centering
\hspace{-.1cm}\includegraphics[scale=.5]{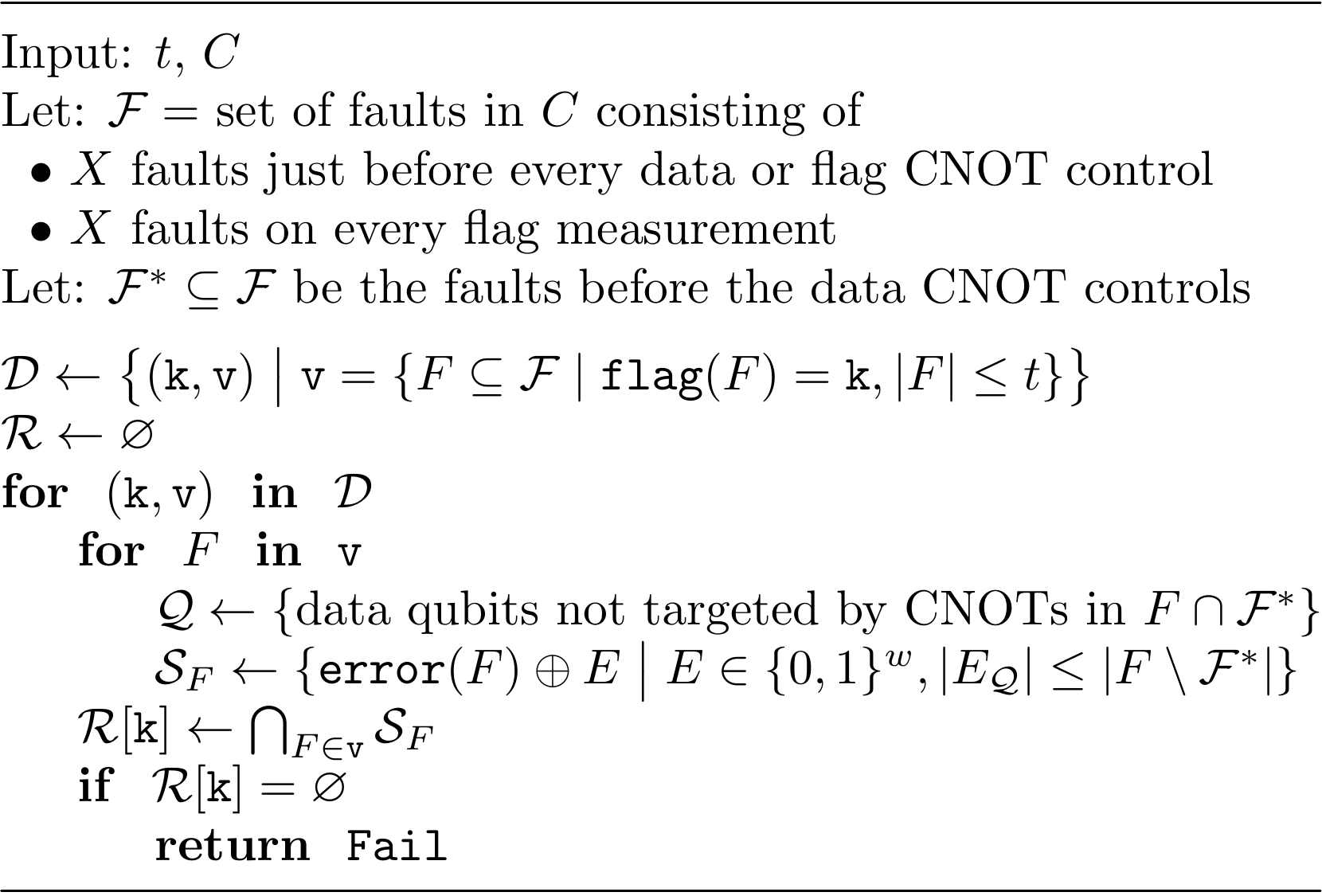}
\end{figure}

\bibliographystyle{alpha-eprint}
\bibliography{q}
\end{document}